\documentclass[twocolumn,conference]{IEEEtran}
\usepackage{amsmath}
\usepackage{amssymb}
\usepackage{color}
\usepackage{bm}
\usepackage[dvipdfmx]{graphicx}
\usepackage{amsmath}
\usepackage{amssymb}
\usepackage{amsfonts}
\usepackage{amsthm}

\newcommand{\RQ}{R_\mathrm{Q}}

\newcommand{\ZP}{{\mathbb Z}_P}

\newcommand{\dc}{n_\mathrm{c}}
\newcommand{\ds}{n_\mathrm{s}}
\newcommand{\dl}{d_{\mathrm{l}}}
\newcommand{\dr}{d_{\mathrm{r}}}
\newcommand{\dt}{d_{\mathrm{t}}}

\newtheorem{theorem}{Theorem}[section]

\newtheorem{example}{Example}[section]
\newtheorem{remark}{Remark}[section]

\begin{document}
\title{Spatially Coupled Quasi-Cyclic Quantum LDPC Codes}
\author{
\authorblockA{
Manabu HAGIWARA${}^{\dagger}$ \quad Kenta KASAI${}^{\ddagger}$ \quad Hideki IMAI${}^{\dagger *}$ \quad Kohichi SAKANIWA${}^{\ddagger}$\\\\
\begin{tabular}{ccc}
\begin{minipage}{5.9cm}
\begin{center}
 ${}^{\ddagger}$Research Center for\\
 Information Security, \\
National Institute of Advanced Industrial Science and Technology, \\
 101-002  Tokyo, JAPAN.\\
\end{center}
\end{minipage} 
&\begin{minipage}{5.5cm}
\begin{center}
   ${}^{\dagger}$Dept.~of Communications \\
and Integrated
 Systems, \\Tokyo Institute of Technology, \\152-8550 Tokyo, JAPAN\\
\end{center}
\end{minipage} 
&\begin{minipage}{5.5cm}
\begin{center}
  ${}^{*}$Dept.~of Electrical, Electronic and Communication Engineering, \\
 Faculty of Science and Engineering, Chuo University, \\
 112-8551 Tokyo, JAPAN.
\end{center}
\end{minipage}
\end{tabular}
\vspace{5mm} \\
Email:{\ttfamily \{hagiwara.hagiwara, ~h-imai\}\allowbreak@aist.go.jp},\hspace{6.6mm} {\ttfamily \{kenta, sakaniwa\}@\allowbreak comm.ss.titech.ac.jp}
}
}
\maketitle

\begin{abstract}
We face the following dilemma for designing low-density parity-check codes (LDPC) for quantum error correction. 
1) The row weights of parity-check should be large: The minimum distances are bounded above by the minimum row weights of parity-check matrices of 
constituent classical codes. Small minimum distance tends to result in poor decoding performance at the error-floor region. 
2) The row weights of parity-check matrices should not be large: The sum-product decoding performance at the water-fall region is degraded 
as the row weight increases.\\
Recently, Kudekar et al.~showed spatially-coupled (SC) LDPC codes exhibit capacity-achieving performance for classical channels. 
SC LDPC codes have both large row weight and capacity-achieving error-floor and water-fall performance. 
In this paper, we design SC LDPC-CSS (Calderbank, Shor and Steane) codes for quantum error correction over the depolarizing channels. 
\end{abstract}


\section{Introduction}
In 1963, Gallager invented low-density parity-check (LDPC) codes \cite{gallager_LDPC}, which is defined as a kernel of a sparse parity-check matrix.
Due to the sparseness of the parity-check matrix, 
LDPC codes are efficiently decoded by the sum-product (SP) algorithm. 
messages of SP decoding can be statistically evaluated. 
Since the optimized LDPC codes can approach very close to the Shannon limit \cite{richardson01design},
error-correcting code theorists are attracted by LDPC codes.

By the discovery of CSS (Calderbank, Shor and Steane) codes \cite{calderbank96, steane96b} and stabilizer codes \cite{gottesman96},
the notion of parity-check measurement is introduced as a generalized notion of parity-check matrix.
From a point of view of this generalization, quantum LDPC codes are naturally defined via low-density parity-check measurements \cite{Pos01a} by Postol.
In particular, a parity-check measurement for a CSS code is characterized by a pair of parity-check matrices.
If both of these parity-check matrices of the pair are sparse parity-check matrix, the related CSS code is called a quantum LDPC (CSS) code.

For classical case, a random construction method for constructing a sparse parity-check matrix generate a high error-correcting performance LDPC code.
On the other hand, it is almost impossible to apply the same method for quantum case, since
the pair of parity-check matrices have to satisfy the following constraint: the product of one of the pair and the transposed other is subjected to be a zero-matrix.
Therefore one of the research interests for quantum LDPC codes is stated as: ``Achieve the constraint and sparseness simultaneously.
Additionally, valuable structure for classical LDPC codes is also achieved.''

MacKay et al. proposed the {\it bicycle} codes \cite{1337106} and Cayley graph based CSS codes \cite{MoreSparseGraph}.
These codes are known as \textit{self-dual containing} LDPC codes.
In \cite{4557323}, two of the authors proposed a construction method of CSS code pair that has \textit{quasi-cyclic (QC)} parity-check matrices with
\textit{arbitrary regular} even row weight $\dr\ge 4$ and column weight $\dl$ such that $\dr/2\ge \dl\ge 2$. 
In \cite{KHIK10}, the authors generalized this construction method to codes over \textit{non-prime field of characteristic number 2}. 
To the best of the authors' knowledge, these codes  \cite{1337106, MoreSparseGraph, 4957637} hold the highest error-correcting performance
among efficiently decodable quantum LDPC codes so far. 

Spatially-coupled (SC) LDPC codes are classical capacity-achieving codes based on the construction of convolutional LDPC codes.
Felstr{\"o}m and Zigangirov \cite{zigangirov99} introduced a construction method of $(\dl, \dr)$-regular convolutional LDPC codes from $(\dl, \dr)$-regular block LDPC codes. 
Surprisingly, the LDPC convolutional exhibited better decoding performance than the underlying block LDPC codes under a fair comparison with respect to the code length. 
Note that in this paper,  convolutional LDPC codes are the LDPC codes defined by band sparse parity-check matrices. 

Kudekar et al.~ named this phenomenon as ``threshold saturation'' and proved rigorously for the binary erasure channels \cite{DBLP:journals/corr/abs-1001-1826}. 
Further,  Kudekar et al.~ \cite{DBLP:journals/corr/abs-1004-3742} 
observed empirical evidence which support the threshold saturation occurs also for the binary-input memoryless symmetric-output (BMS) channels.
Another remarkable advantage of SC LDPC codes is encoder universality.
In other words,  keeping $\dl / \dr$, in the limit of large $\dl$, $L$ and $w$,  the coupled ensemble $(\dl,\dr,L,w)$ \cite{959254}
achieves {\em universally} the capacity of the BMS channels under SP decoding. 
Conventional capacity-achieving codes such as  polar codes \cite{5075875} and irregular LDPC codes \cite{richardson01design} do not support such universality.
According to the channel, polar codes and need selection of frozen bits \cite{DBLP:journals/corr/abs-0901-2207}
and irregular LDPC codes need optimization of degree distributions. 

In this paper, we propose a construction method for \textit{spatially coupled quantum LDPC codes} for quantum error correction.
By the proposed method, we obtain a pair of parity-check matrices such that these matrices are each orthogonal and each matrix has a diagonal band structure.

\section{Preliminaries}
Let us recall definitions of related fundamental notion of quantum LDPC codes
 and classical SC LDPC codes.
\subsection{CSS codes}
A \textbf{CSS code} $Q$, which is the main interest of this paper, is a class of quantum error codes.
The code is a complex vector space.
The vector space $Q$ is characterized by a pair of classical binary linear codes $C$ and $D$ whose parity-check matrices $H_C$ and $H_D$ satisfy $H_C H_D^\mathsf{T}=0$.

A CSS code $Q$ associated with $(H_C, H_D)$ is defined as a complex linear combination of the following vectors:
$$ \sum_{d' \in D^{\perp}} | c + d' \rangle \;\; \text{ for } c \in C,$$
where $D^{\perp}$ is the dual code of $D$ as a classical code
 and the basis for quantum states is assumed as a standard computational basis.

\subsection{LDPC-CSS codes}
When the parity-check matrices $H_C$ and $H_D$ are sparse, 
the CSS code associated with $(H_C, H_D)$ are called an \textbf{LDPC-CSS code}. 
As it is written above, the structure of the code space $Q$ is completely characterized by two linear codes $C$ and $D$ associated with $(H_C, H_D)$.
The aim of this paper is to construct these codes as LDPC-CSS codes, particularly SC LDPC codes.

One of issues for designing $(H_C, H_D)$ is the row weights of these matrices.
LDPC-CSS codes are efficiently decoded by using  SP syndrome decoding \cite{1337106}. 
Its success probability for decoding is highly affected by the row weight of these parity-check matrices $(H_C, H_D)$:
\begin{enumerate}
\item The row weights of each of $H_C$ and $H_D$ should be large: 
The minimum distances of $C$ and $D$ are less than or equal to the minimum row weights of $H_D$ and $H_C$, respectively. 
Small minimum distance tends to result in poor decoding performance at the error-floor region. 
\item The row weights of each of $H_C$ and $H_D$ should not be large: The SP decoding performance at the water-fall region are degraded if row weight of $H_D$ and $H_C$ increase.
\end{enumerate}
In summery, we are facing the dilemma above for designing $(H_C, H_D)$.
The proposed idea to overcome the dilemma is
 to introduce  a class of classical ``spatially coupled LDPC codes'' to quantum LDPC codes.

\subsection{Sparse Band Model Matrices and Spatially Coupled LDPC Codes}
Here, we construct classical SC LDPC codes from \textbf{sparse band model matrices} $\mathcal{H}$.
Let $\dl, \dt$ and $\dc$ be positive integers.
The integer $\dl$ shall be the column weight of $\mathcal{H}$.
Let $\ds$ be a positive integer such that $\ds$ divides $\dl$.
Briefly speaking on the structure of $\mathcal{H}$,
submatrices of size $\dl \times \dt$ appear $\dc$ times on the diagonal line.
Thus the column weight of $\mathcal{H}$ is $\dl$.
Define $\dr := \dt \dl/\ds$.
The row weight of $H$ is $\dr$ in the middle and $\dt$ in the top and bottom. 
Define $M := \dl + (\dc-1)\ds$ and $N := \dc \dt$.
The size of $\mathcal{H}$ is $M \times N$.

We give a formal construction for the \textbf{sparse band model matrix}.
Define a matrix $\mathcal{H} = (h_{j,l})$ over $\{0, * \}$ of size $(\dl + (\dc-1)\ds) \times \dc \dt$
 by putting
$h_{j, l} := *$ if $ b\ds+1 \le j \le (b+1)\ds$ or $h_{j, l} = 0$, where $*$ is a formal symbol and $ b := \lfloor (l-1)/\dt \rfloor$.

The following is an example of a sparse band model matrix $\mathcal{H}$ for parameters $\dl=4, \dt=2, \dc=12$ and $\ds=1$.
\begin{center}
$\mathcal{H}= \left[
  \begin{minipage}{5.1cm}
 \vspace{1mm} 
 \renewcommand{\baselinestretch}{0.3}
 \begin{verbatim}
**0000000000000000000000
****00000000000000000000
******000000000000000000
********0000000000000000
00********00000000000000
0000********000000000000
000000********0000000000
00000000********00000000
0000000000********000000
000000000000********0000
00000000000000********00
0000000000000000********
000000000000000000******
00000000000000000000****
0000000000000000000000**
 \end{verbatim}
 \renewcommand{\baselinestretch}{1.0}
  \end{minipage}
 \right].$
\end{center}
The sparse band model matrix $\mathcal{H}$ contains $\dc (=12)$-subblocks of size $\dl (=4) \times \dt (=2)$.

Next, we define a $(\dl,\dr,\dc)$ SC LDPC code via the model matrix $\mathcal{H}$. 
Let $P$ be a positive integer.
Replace each ``*'' of $\mathcal{H}$ with a binary permutation matrix of size $P\times P$
and each $0$ of $\mathcal{H}$ with a zero matrix of size $P \times P$.
The replaced matrix is called \text{spatially couple LDPC matrix} and its kernel space is called a classical \textbf{spatially coupled LDPC code}.
For fixed $\dr$ and $\dl$, the size $PM\times PN$ of such a sparse band matrix scales as $M=N\dl/\dr+\dl-1$ increases.
The value $R:=1-M/N$ is called a \text{design rate}.
The design rate converges to $1-\dl/\dr$ as $\dc$ tends to infinity.
The SC LDPC code ensemble  \cite{DBLP:journals/corr/abs-1001-1826} was shown to exhibit SP decoding performance which is very close to the MAP decoding performance of 
$(\dl,\dt)$-regular LDPC code ensemble \cite{mct} in the limit of large code length and coupling number $\dc$. 
For a sparse band matrix $\mathcal{H}$,
 define a matrix $\mathcal{H}' = (h'_{j, l})$ over $\{ 0, * \}$
 by putting $h'_{j, l} := h_{ r - j +1 , l}$, where $r$ is the number of rows of $\mathcal{H}$.
We also call $\mathcal{H}'$ a sparse band model matrix.

SC LDPC codes enable to have large minimum row weight of parity-check matrix and simultaneously exhibit excellent SP decoding performance. 
Hence, in this paper, we study the design of SC LDPC-CSS codes. It is expected that such codes overcome the dilemma on the row weight. 
\subsection{Quasi-Cyclic LDPC Codes}
In the previous works, the decoding performance of a SC LDPC code
 is analyzed by the technique based on ensemble analysis.
In other words, the instance of SC LDPC code shall be constructed from
randomly chosen permutation matrices for a sparse band model matrix.
On the other hand, it is not expected that the requirement for $(H_C, H_D)$,
i.e. $H_C H_D^\mathsf{T} = 0$, holds on parity-check matrices of SC LDPC codes by using random choice. 
Therefore, we slightly relax the constraint of spatially-coupled LDPC codes.
We employ circulant permutation matrices instead of permutation matrices. 

Let us define a matrix $I(1)$ over binary field of size $P \times P$ by putting:
\begin{align*}
 I(1) :=
  \begin{bmatrix}
   0 & 1 & 0 & 0 & 0 \\
   0 & 0 & 1 & 0 & 0 \\
   0 & 0 & 0 & \ddots & 0 \\
   0 & 0 & 0 & 0 & 1 \\
   1 & 0 & 0 & 0 & 0   
  \end{bmatrix}\in \{0, 1\}^{P\times P}.
\end{align*}
In other words, $I(1)$ is a circulant permutation matrix which is a single right shifted matrix for the identity matrix.
For an integer $x$, define a $P \times P$ matrix $I(x) :=I(1)^{x}$.
$I(0)$ is the identity matrix.
For a symbol $\infty$, define a $P \times P$ matrix $I(\infty)$ as a zero matrix.

For a matrix $(c_{j, l})_{0 \le j < d_1, 0 \le l < d_r}$ over $\{0, 1, \dots, P-1, \infty \}$,
let $H$ be a $P\dl\times P\dr$ binary parity-check matrix defined as follows:
 \begin{align*}
 &{H} := (I(c_{j, \ell}))_{0\le j<\dl, 0\le \ell<\dr}. \\
\end{align*}
We refer to such a matrix as \textbf{($\dl, \dr, P$) quasi-cyclic (QC) permutation matrix},
 or simply a \textbf{QC permutation matrix}.
The associated linear code is called a \textbf{QC LDPC code}.
\section{Spatially Coupled LDPC-CSS Codes based on Quasi-Cyclic Permutation Matrices}
\label{230852_10Jun10}
Let $(H_C, H_D)$ be an LDPC-CSS code.
If $H_C$ and $H_D$ are SC LDPC matrices,
 we call the LDPC-CSS code a \textbf{spatially coupled LDPC-CSS code}.
It is known that small cycles in the Tanner graph degrade the SP decoding performance.
In this section,
 we construct SC LDPC matrix pair $(H_C,H_D)$
 which satisfies the following three conditions.
\begin{enumerate}
\setlength{\itemsep}{-0mm}
 \item $H_C H_D^{\mathsf{T}}=0$
 \item Both of the Tanner graphs of $H_C$ and $H_D$ are free of cycles of length 4,
 \item $H_C$ and $H_D$ are QC-permutation matrices,
\end{enumerate}

\subsection{Conventional QC-LDPC CSS codes}
We first review the construction developed in \cite{4557323}.
The authors proposed \cite{4557323} the following method for constructing a QC parity-check matrix pair $({H}_C, {H}_D)$. 
In the original paper \cite{4557323}, the construction method is flexible about the row size of the matrices, i.e., ${H}_C$ and ${H}_D$ can have different row sizes. 
For simplicity, in this paper,  we focus our attention on the case that ${H}_C$ and ${H}_D$ have the same row weight $\dr$ and have the same column weight $\dl$.

Let $(c_{j, l})$ and $(d_{k, l})$ be matrices over $\{0, 1, \dots, P-1 \} \cup \{ \infty \}$ of size $\dl \times \dr$.
Let $H_C$ (resp. $H_D$) be a QC-permutation matrix associated with $(c_{j,l})$ (resp. $(d_{k, l})$).
From \cite{4557323}, it is proved that
 ${H}_C{H}_D^{\mathrm{T}}=0$, if $\#\{\ell\in [0,P-1]\mid c_{j,\ell}-d_{k,\ell}=p \mod P\}$ is even for all $ j\in[0,\dl-1]$ and for all  $p \in [0,P-1]$.
The Tanner graph of ${H}_C$ is free of cycles of length 4, if 
\begin{align}
 \#\{c_{j,\ell}-c_{k,\ell} \in \mathbb{Z}_P^* \mid \ell\in [0,\dr-1]\}=\dr\label{232617_3Feb11}
\end{align}
for all  $p\in[0,P-1]$, and for all $j, k \in[0,\dl-1]$ such that $j \neq k$.
We give the following theorem which is slightly generalized version of Theorem 6.1 in \cite{4557323}. 

\begin{theorem}
\label{Hagiwara-Imai}
Let $P$ be a positive integer with $P > 2$.
Define 
 $\ZP^{*} := \{ z \in \ZP\mid \exists a \in \ZP, za=1 \}$.
For $\sigma, \tau \in \mathbb{Z}_P^*$, define
 $\mathrm{ord}(\sigma) := \min\{ m > 0 \mid \sigma^m = 1\}$,
and define
${\langle \tau \rangle}_{\sigma}=\{\tau,\tau\sigma,\dotsc,\tau\sigma^{\mathrm{ord}(\sigma)-1}\}.$

Let $\dl, \dr$ be integers and $\tau_1, \tau_2$ be in $\ZP^{*}$ such that 
\begin{align}
& \dl \ge 2, \dr \ge 4, \nonumber\\
& \dr/2 =  \mathrm{ord}(\sigma), \label{231740_4Jul10}\\
& 1 \le \dl \le \mathrm{ord}(\sigma),\nonumber\\
&\mathrm{ord}(\sigma) \neq \# \ZP^{*},\nonumber \\
&1- \sigma^{j} \in \ZP^{*} \text{ for all } 1 \le j < \mathrm{ord}(\sigma), \nonumber\\
& \tau_2 \not \in \langle\sigma\rangle_{\tau_1}. \label{215258_7Jul10} 
\end{align}
Let ${H}_C$ and ${H}_D$ be two $(\dl, \dr, P)$-QC binary matrices such that 
\begin{align*}
&{H}_C=(I(c_{j, \ell}))_{0\le j<\dl, 0\le\ell<\dr}, \nonumber\\
&{H}_D=(I(d_{j, \ell}))_{0\le j<\dl, 0\le\ell<\dr}, \nonumber
\end{align*}
where
\begin{align*}
&   c_{j, \ell} := 
 \left\{
  \begin{array}{rc}
   \tau_1\sigma^{-j + \ell}     &  0 \le \ell < \dr/2 \\
   \tau_2\sigma^{-j + \ell}     &  \dr/2 \le \ell < \dr,  \\
  \end{array}
 \right.\\
 & d_{j, \ell} := 
 \left\{
  \begin{array}{rc}
    -\tau_2 \sigma^{j-\ell}     &  0 \le \ell < \dr/2\label{051527_15Jun10}  \\
    -\tau_1 \sigma^{j-\ell}     &  \dr/2 \le \ell < \dr.  \\
  \end{array}
 \right.
\end{align*} 
Then, it holds that 
1) $H_C H_D^{\mathsf{T}}=0$,
2) both of the Tanner graphs of $H_C$ and $H_D$ are free of cycles of length 4.
\end{theorem}
\textbf{See \S\ref{proof_for_construction} for the construction proof.}

If $\tau_1 = 1$, the statement is the same as the one in \cite{4557323}.
From Theorem \ref{Hagiwara-Imai}, we obtain two $\dl P\times \dr P$ binary matrices ${H}_C$ and ${H}_D$
such that ${H}_C{H}_D^{\mathsf{T}}=0$ and the Tanner graphs of ${H}_C$ and ${H}_D$ are free of cycles of size 4. 
In this paper, we refer these codes  as conventional QC-LDPC CSS codes. 
The following is an example:
\begin{example}
\label{ex1}
 With parameters $\dl=3, \dr=6, P=7, \sigma=2$ and $\tau_1=1, \tau_2=3$, 
 from Theorem \ref{Hagiwara-Imai}, we obtain a $J P\times \dr P$ binary matrix pair $({H}_C, {H}_D)$ such that ${H}_C{H}_D^{\mathsf{T}}=0$ as follows. 
 \begin{align*}
 & {H}_C=
 \begin{pmatrix}
 I(1)&I(2)&I(4)&I(3)&I(6)&I(5)\\
 I(4)&I(1)&I(2)&I(5)&I(3)&I(6)\\
 I(2)&I(4)&I(1)&I(6)&I(6)&I(3)
 \end{pmatrix}, \\
 &{H}_D=
 \begin{pmatrix}
 I(4)&I(2)&I(1)&I(6)&I(3)&I(5)\\
 I(1)&I(4)&I(2)&I(5)&I(6)&I(3)\\
 I(2)&I(1)&I(4)&I(3)&I(5)&I(6)
 \end{pmatrix}.
 \end{align*}
\end{example}
\subsection{Spatially coupled CSS codes}
\begin{figure*}
\input{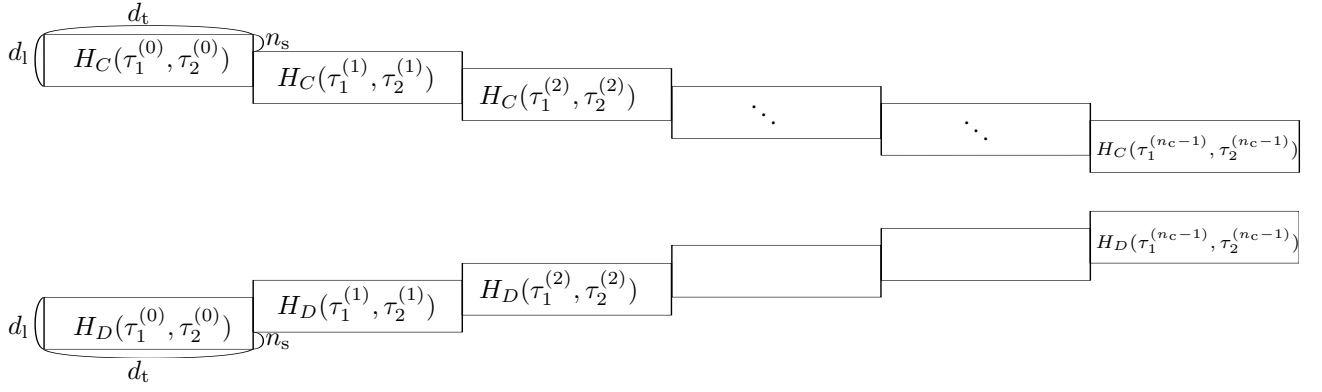}
\caption{Illustration of parity-check matrices $H_C$ and $H_D$ of a spatially-coupled LDPC-CSS code (C,D). }
\label{051353_3Feb11}
\end{figure*}

Let ${H}_C(\tau_1,\tau_2)$ and ${H}_D(\tau_1,\tau_2)$ denote a pair of parity-check matrices associated with $\tau_1$ and $\tau_2$ in Theorem \ref{Hagiwara-Imai}.
Let $\dc$ pairs of matrices ${H}_C(\tau_1^{(i)},\tau_2^{(i)})$ and ${H}_D(\tau_1^{(i)},\tau_2^{(i)})$ for $i=0,\dotsc,\dc-1$.
Let $\ds$ be a positive number such that $\ds|\dl$. 
We construct sparse band QC matrix pair $H_C$ and $H_D$ as depicted in Fig.~\ref{051353_3Feb11}.
It holds that  $H_C H_D^\mathsf{T}=0$ since $H_C(\tau_1^{(i)},\tau_2^{(i)})H_D(\tau_1^{(i)},\tau_2^{(i)})^\mathsf{T}=0$  for $i=0,\dotsc,\dc-1$.
Therefore the orthogonality $H_C H_D^\mathsf{T}=0$ does not depend on the choice of $\tau_1^{(i)}, \tau_2^{(i)}$.
$H_C$ and $H_D$ have regular-column weight $\dl$ and slightly irregular row weight. The row weight
 is typically $\dt\dl/\ds$ around the center of $H_C$ and $H_D$ and takes the minimum weight $\dt$ at the boundaries.

On the other hand, the condition ``there are no cycles of size 4 in the Tanner graph'' depends on the choice of them,
since the $H_C(\tau_1^{(i)},\tau_2^{(i)})$ and $H_C(\tau_1^{(i')},\tau_2^{(i')})$ share a common row if $|i- i'|< \dl / \ds$.
By a similar argument in the proof of Theorem \ref{Hagiwara-Imai},
we have the following theorem:
\begin{theorem}
If $\langle \tau_b^{i} \rangle_{\sigma} \cap \langle \tau_b'^{i'} \rangle_{\sigma} = \emptyset $
for any $b, b' \in \{ 1, 2 \}$ and $i, i' \in \{ 0, 1, \dots, \dc -1 \}$ such that $|i - i' | < \dl/\ds$,
then ${H}_{C}{H}_D^{\mathsf{T}} = 0$ and there are no cycles of size 4 
in the the Tanner graphs of sparse band QC matrices ${H}_C$ and ${H}_D$. 
\end{theorem}
\begin{example}
 In Fig.~\ref{090405_9Feb11}, we give an example of ${H}_C$ and ${H}_D$ of SC LDPC-CSS codes with 
$\dl=3, \dt=6, P=31, \dc=6, \ds=1,
(\tau_0^{(0)},\tau_1^{(0)})=(16,4),\allowbreak 
(\tau_0^{(1)},\tau_1^{(1)})=(8,12), \allowbreak 
(\tau_0^{(2)},\tau_1^{(2)})=(6,1), \allowbreak 
(\tau_0^{(3)},\tau_1^{(3)})=(3,11), \allowbreak 
(\tau_0^{(4)},\tau_1^{(4)})=(17,2), \allowbreak 
(\tau_0^{(5)},\tau_1^{(5)})=(6,4). 
$\end{example}
\begin{figure*}
\input{example_CSS_SC}
\caption{
Parity-check matrices ${H}_C, {H}_D$ of the proposed SC LDPC-CSS codes defined by $\dl=3, \dt=6, P=31, \dc=6, \ds=1$. $I(\cdot)$ is abbreviated for each entry.
}
\label{090405_9Feb11}
\end{figure*}
\begin{remark}
The Tanner graphs for $H_C$ and $H_D$ are isomorphic.
This statement is proved by combinatorial argument
although we omit the proof in this paper.
\end{remark}

\section{Numerical Results}
We assume the transmission takes place over depolarizing channels \cite[Section V]{1337106} with depolarizing probability $2f_\mathrm{m}/3$, where $f_\mathrm{m}$ can be viewed as the marginal probability for 
$\mathtt{X}$ and $\mathtt{Z}$ errors. 
Note that the channel is the normal depolarizing channel. 
We assume the decoder knows the depolarizing probability $3f_\mathrm{m}/2$. 

\begin{figure}
\includegraphics[width=0.48\textwidth]{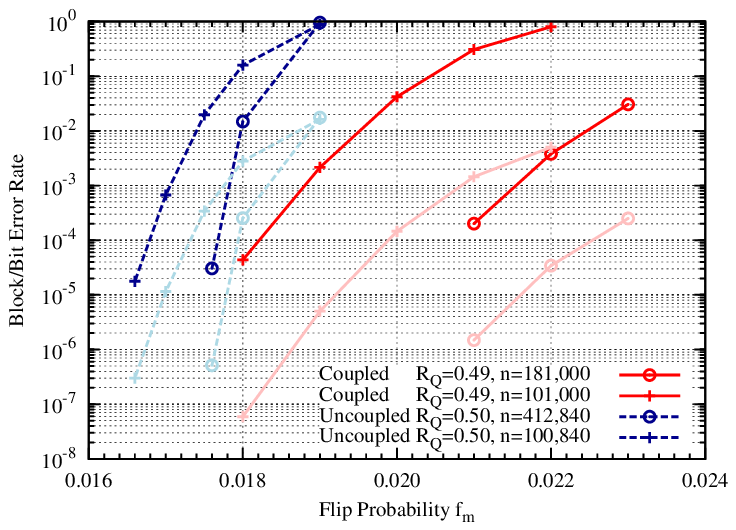} 
\caption{
Comparison of decoding error rate of the proposed SC LDPC-CSS code (red) with ($\dl=10,\dt=20,\dc=50,\ds=5$) and conventional QC-LDPC CSS codes (blue) with ($\dl=10,\dr=40$). 
The quantum coding rate is $\RQ=0.49$ and $\RQ=0.50$, respectively. The code length is $n$ qubits.
The block (dark) and bit (light) error rate of the constituent codes $C$ and $D$ of 
the proposed CSS code pair $(C, D)$ over the depolarizing channel with marginal flip probability $f_\mathrm{m}$ of $\mathtt{X}$ and $\mathtt{Z}$ errors. 
Due to the isomorphism, the error rates of $C$ and $D$ are the same. 
}
\label{073259_14Feb11}
\end{figure}

Figure \ref{073259_14Feb11} compares the proposed SC LDPC-CSS codes and conventional QC-LDPC CSS codes.
The quantum coding rate of these codes is 0.50 and 0.49, respectively.  
Such rate-loss is due to the coupled construction. The rate-loss can be reduced by increasing $\dc$.
The parity-check matrices of these codes have the same column weight 10. The minimum row weight 40 and 20, respectively. 
For both proposed and conventional CSS codes, two component classical codes $C$ and $D$ are isomorphic. 
This isomorphism implies that $C$ and $D$ have the same average decoding performance. 
Hence, we restrict our attention only to the performance of $C$. 

Due to the large minimum row weight of parity-check matrix, all the codes exhibit no error-floors at the bit error rate down to $10^{-6}$. 
No undetected errors were observed for both codes. 
Hence, it is expected the minimum distances of both codes are sufficiently large and tight up to the upper bounds 40 and 20, respectively.
As posed in the dilemma, it is observed that conventional QC-LDPC CSS codes have only a small coding gain by increasing code length from $n=100840$ to $n=412840$.
The proposed code outperforms the conventional code with much shorter code length $n=101000$ and $n=181000$. 
This excellent performance outweighs the rate-loss.  
\section{Conclusion}
In this paper, we proposed spatially-coupled LDPC-CSS codes.  
Numerical experiments show that the proposed codes have both deep error-floors and excellent water-fall performance over the depolarizing channels. 
In other words, threshold saturation is also observed for the LDPC-CSS codes. 
\bibliographystyle{IEEEtran}
\bibliography{IEEEabrv,kenta_bib}
\section{Appendix}\label{proof_for_construction}
\begin{proof}[Proof for SC Construction (Theorem \ref{Hagiwara-Imai})]
1)
Note that 
\begin{align*}
&\{c_{j,\ell}-c_{j' ,\ell}\in\ZP\mid 0\le\ell<\frac{\dr}{2}\}
 ={\langle \tau_1(\sigma^{-j}-\sigma^{-j'}) \rangle}_{\sigma},\\
&\{c_{j,\ell}-c_{j' ,\ell}\in\ZP\mid \frac{\dr}{2}\le\ell<\dr\}
 ={\langle \tau_2(\sigma^{-j}-\sigma^{-j'}) \rangle}_{\sigma},
\end{align*}
where $j, j' \in[0,\dl-1]$ such that $j \neq j'$.
Therefore 
$$ \# \{c_{j , \ell}-c_{j' , \ell}\in\ZP\mid 0\le\ell<\frac{\dr}{2}\} = \dr / 2$$
and
$$ \# \{c_{j,\ell}-c_{j' ,\ell}\in\ZP\mid \frac{\dr}{2}\le\ell<\dr\} = \dr / 2.$$
In general,
$$ \langle x \rangle_{\sigma} \cap \langle x' \rangle_{\sigma} = \emptyset
\iff
\langle x y \rangle_{\sigma} \cap \langle x' y \rangle_{\sigma} = \emptyset
$$
for any $x, x', y \in \mathbb{Z}_P^*$.
By \eqref{215258_7Jul10}, 
$$ \langle \tau_1 \rangle_{ \sigma } \cap \langle \tau_2 \rangle_{ \sigma } = \emptyset.$$
It implies that 
\begin{align*}
& {\langle \tau_1(\sigma^{-j}-\sigma^{- j' }) \rangle}_{\sigma}\cap{\langle \tau_2(\sigma^{-j}-\sigma^{-j' }) \rangle}_{\sigma}=\emptyset
\\
&\iff
 {\langle \tau_1 \rangle}_{\sigma}\cap{\langle \tau_2 \rangle}_{\sigma}=\emptyset.
\end{align*}
Hence
\begin{eqnarray*}
& &
 \# \{c_{j,\ell}-c_{j' ,\ell}\in\ZP\mid 0\le\ell< \dr\} \\
& = &
\# \{c_{j,\ell}-c_{j' ,\ell}\in\ZP\mid 0\le\ell<\frac{\dr}{2}\}\\
&  & + \; 
\# \{c_{j,\ell}-c_{j' ,\ell}\in\ZP\mid \frac{\dr}{2}\le\ell<\dr\} \\
& = & \dr /2 + \dr /2 = \dr. 
\end{eqnarray*}
By \eqref{232617_3Feb11}, there are no cycles of size 4 
in the the Tanner graph of ${H}_C$. We can show for $H_D$ in a similar way. 

2)
By direct calculation,
\begin{eqnarray*}
& & \{c_{j,\ell}-d_{k,\ell}\in\ZP\mid 0\le\ell<\frac{\dr}{2}\}\\
& = & \{c_{j,\ell}-d_{k,\ell}\in\ZP\mid \frac{\dr}{2}\le\ell<\dr\}.
\end{eqnarray*}
Therefore 
${H}_C{H}_D^{\mathrm{T}}=0$, if $\#\{\ell\in [0,P-1]\mid c_{j,\ell}-d_{k,\ell}=p \mod P\}$ is even for all $ j\in[0,\dl-1]$ and for all  $p \in [0,P-1]$.
It means $H_C H_D^\mathsf{T} = 0$.
\end{proof}

\end{document}